
{\magnification=\magstep1
\def\refs{\leftskip=.3truein\parindent=-.3truein}
\def\endrefs{\leftskip=-.3truein\parindent=.3truein}
\baselineskip=24pt

\centerline{ENERGY PRINCIPLES FOR SELF-GRAVITATING BAROTROPIC FLOWS:}
\centerline{II. THE STABILITY OF MACLAURIN FLOWS}

\bigskip

\centerline{Asher Yahalom and Joseph Katz}

\centerline{The Racah Institue of Physics, Jerusalem 91904, Israel}

\centerline{and}

\centerline{Shogo Inagaki}

\centerline{Department of Astronomy, Faculty of Science, Kyoto University,}

\centerline{Sakyo-ku, Kyoto 606-01}

\bigskip
\noindent
Submitted for Publication to M.N.R.A.S (July 1993). Figures will be sent to the
reader upon request. (contains 36 pages)

\vfill\eject

\centerline{\bf Abstract}
\noindent
	We analyze stability conditions of "Maclaurin flows" (self-gravitating,
barotropic, two dimensional, stationary
 streams moving in closed loops around a point) by minimizing their  energy,
subject to fixing all the constants of the motion including mass and
circulations. Necessary and sufficient conditions of stability are obtained
when gyroscopic terms in the perturbed Lagrangian are zero. To illustrate and
check the properties of this new energy principle, we have calculated the
stability limits of an ordinary Maclaurin disk whose dynamical stability limits
are known. Perturbations are in the plane of the disk. We find all necessary
and sufficient conditions of stability for single mode symmetrical or
antisymmetrical perturbations. The limits of stability are identical with those
given by a dynamical analysis. Regarding mixed types of perturbations the
maximally constrained energy principle give for some the necessary and
sufficient condition of stability, for others only sufficient conditions of
stability. The application of the new energy principle to Maclaurin disks shows
the method to be  as powerful as the method of dynamical perturbations.

\bigskip

\vfill\eject

\noindent
{\bf 1. Introduction}
\bigskip
	The main purpose of this work is to illustrate on a test case our principle of
"maximally constrained" energy minimum (Katz, Inagaki and Yahalom 1993), and
find stability limits of stationnary motions of barotropic flows. Paper I dealt
with three dimensional flows and gave no application. Here we consider two
dimensional "Maclaurin Flows" and apply the method to Maclaurin disks.  With
"Maclaurin Flows" we mean some sort of generalized rotating streams, with
closed loops around a point, but no special symmetry. The simplest and most
symmetric example of a Maclaurin flow is a Maclaurin disk. Other two
dimensional fluid models for galaxies are Maclaurin flows as well. Stationary
flows are "stationary points" of their energy. Flows are stable if their energy
is minimum, compared to that of all slightly different stationary trial
configurations . By constraining trial configurations to satisfy all the
constraints that a dynamical flow would satisfy: linear momentum, angular
momentum, mass conservation and circulation, we may  greatly enhance the value
of the energy principle. Energy principles give only sufficient conditions of
stability. However, we know that if the "gyroscopic term" in the Lagrangian of
a dynamically perturbed configuration is zero, a minimum of energy becomes a
necessary and sufficient condition f

 Our test model is a Maclaurin disk. This is one of  those rare models (Binney
\& Tremaine 1987) were a detailed dynamical stability analysis is possible. It
is thus the one valuable test case in which the details of our method and its
results can be compared with a different, independant calculation. Both method
use spherical harmonic decompositions of perturbations or $(l,m)$ - modes.

What we find are the { \it same stability limits} for all symmetrical and
antisymmetrical single modes as those given by the  dynamical analysis of
Binney and Tremaine. Other models have been tested which give also necessary
and sufficient conditions of stability. Thus, Yahalom (1993) found the
stability limits of a uniform rotating sheet and of two dimensional Raleigh
flows, by the present technique, to be the same as that of a dynamical
analysis. The new energy principle appears thus as a powerful instrument that
may complete and occasionally  replace the method of linear perturbation
analysis. This is the main point of this work.
\bigskip

\noindent
{\bf 2. Description of Stationary Maclaurin Flows, and of the Energy Principle
for their Stability. }
\bigskip

\noindent
{\it 2.1.  Lagrange Variables and Equations of Motion for Stationary Flows.}

	Equations of motion in three and two dimensions are
similar, though in two dimensions angular velocity, angular-momentum and
vorticity have only one component, in the z direction, and there are less
variables. We use the following notations: for the positions of fluid
elements $ \vec R $ or $ (x^K)=(x,y);  K,L,... = 1,2,$ the density of matter
per unit
surface is $ \sigma $; $ \vec W$ is the velocity field in inertial coordinates
and $ \vec U$ the
velocity relative to coordinates moving with uniform velocity $ \vec b$  and
rotating with constant angular velocity $ \Omega_c $ ; these are the
coordinates in which configurations appear stationary. Thus, by definition,
$$ \vec W = \vec U + \vec \eta_c, \eqno(2.1.a)$$
$$ \vec \eta_c = \vec b + \vec \Omega_c \times \vec R \eqno(2.1.b)$$
$$ \vec \Omega_c = \vec 1_z \Omega_c. \eqno(2.1.c)$$
The mass is conserved in moving coordinates
$$ \vec \nabla \cdot (\sigma \vec U) = 0 \eqno(2.2)$$
On the boundary of free self-gravitating fluids the density and the pressure
are zero
$$ \qquad \sigma|_B = 0, \eqno(2.3.a)$$
$$  \qquad P|_B = 0 \eqno(2.3.b)$$
and in flows of astrophysical interest the velocity of sound
is there zero also,
$$ {\partial P \over \partial \sigma }|_{\sigma = 0} = 0. \eqno(2.4)$$
Notice that (2.2) and (2.3) imply
$$ (\vec U \cdot \vec \nabla \sigma)|_B = 0. \eqno(2.5)$$
Since $ \vec \nabla \sigma|_B $ is normal to the boundary as follows from
(2.3a), (2.5) means, as one expects, that $ \vec U $ is tangent to
the boundary. Because of (2.3.a) an integral of a divergence of the form $ \vec
\nabla \cdot ( \sigma \vec F)$ must thus also be
zero by virtue of Green's identity:
$$ \int_S \vec \nabla \cdot (\sigma \vec F) d^2 x = \oint_B \sigma \vec F \cdot
d \vec B = 0 \eqno(2.6)$$
provided $ \vec F$ is single valued and continuous in $S$ and on $B$. Euler's
equations for stationary motions in moving coordinates are well known and can
be written:
$$ \vec U \cdot \vec \nabla \vec W + \vec \Omega_c \times \vec W + \vec \nabla
(h + \Phi) = 0 \eqno(2.7)$$
where the specific enthalpy
$$ h = \int {dP \over \sigma} = \varepsilon(\sigma) - {P \over \sigma},
\eqno(2.8)$$
$\varepsilon$ is the specific internal  energy. The
gravitational potential $\Phi $ is, in the plane of the motion, given by
$$ \Phi (\vec R) = -G \int {\sigma (\vec R') \over |\vec R - \vec R'|} d^2 x.
\eqno(2.9)$$
The vorticity of the flow
$$ \vec \omega = {\rm rot}~ \vec W = {\rm rot}~ \vec U + 2 \vec \Omega_c
\eqno(2.10.a)$$
$$ \vec \omega = \omega \vec 1_z \eqno(2.10.b)$$
\bigskip

\noindent
{\it 2.2. Lagrange Variables for Maclaurin Flows.}
\bigskip
	The labels of fluid elements are appropiate Lagrange variables in fluid
mechanics (Lamb 1945). A continuous labelling of fluid elements depends on the
particular topology of the flow. Here we describe a special labelling, and a
representation of the velocity field for which mass conservation and
circulation conservation hold automatically. We consider Maclaurin flows that
move in
closed loops around a fixed point. Suppose that, in some coordinates moving
with constant velocity $ \vec b$ and rotating with constant angular velocity $
\Omega_c $,  we are given the density $\sigma (x^K)$ and the relative velocity
of the flow $\vec U (x^K)$. We do not suppose that $\sigma$ and $\vec U$
represents a solution of Euler's equations (2.7). We rather regard $\sigma$ and
$ \vec U$ as a {\it trial} configuration of the flow. With $\vec U, \vec b $
and $\Omega_c$  we define a velocity $\vec W (x^K) $ in inertial coordinates by
equation (2.1). Given $\sigma$ and $\vec W$ we can construct a quantity
$$ \lambda = {\sigma \over \omega}\eqno(2.11)$$
where $ \omega $ is the vorticity defined in equation (2.10). If $\sigma|_B =
0$,
$$\lambda|_B = 0 \eqno(2.12)$$
In time dependant flows, the quantity $\lambda $ is conserved along the motion;
$\lambda$ is the inverse
of the potential vorticity; it is also the load of Lynden-Bell and Katz
(1982). The conservation of $\lambda$ is equivalent to the conservation of
circulation
 along closed contours (Katz and Lynden-Bell 1985). Now consider all the closed
loops
of constant $\lambda$ that can be drawn in our trial configuration and
parametrize them with the value of the circulation along each loop $ 2 \pi
\alpha(\lambda) $ ; thus
$$ \alpha (\lambda) = {1 \over 2\pi} C(\lambda) = {1 \over 2 \pi}
\oint_{\lambda} \vec W \cdot d \vec R \eqno(2.13)$$
$ \alpha (\lambda)$ equals zero at the fixed point $O$ of fig. 1 and  along the
boundary $\alpha(0) = \alpha_B$. For simplicity we assumed that $ \alpha$
increases uniformally between $0$ and $\alpha_B$. Having defined $\alpha (x^K)$
at every point of our trial flow, we now define an "angle" $ \beta$  by the
condition that
$$ \vec \omega = \vec \nabla \alpha \times \vec \nabla \beta \eqno(2.14.a)$$
or
$$ \omega = { \partial (\alpha,\beta) \over \partial (x,y)} \eqno(2.14.b)$$
This is a first order differential equation which defines $\beta$ up to an
arbitary function of $\alpha$, $B( \alpha)$ that depends on the choice of the
line $\beta = 0$ (see figure 1). We shall define $\beta = 0$ as the positive x
axis. Following (2.13)
$$ \alpha = {1 \over 2 \pi} \oint_{\lambda} \vec W \cdot d \vec R = {1 \over 2
\pi} \int \omega dx dy = {1 \over 2 \pi} \int d\alpha d\beta = \alpha (\lambda)
{1 \over 2 \pi} \oint_{\lambda}  d \beta \eqno(2.15)$$
The domain of $ \beta $ is thus $ 0 \leq \beta \leq 2\pi$. Having now an
$\alpha (x^K)$ and a $\beta (x^K) $, it follows from (2.10) and (2.14.a) that
the velocity field is of the form
$$ \vec W = \alpha \vec \nabla \beta + \vec \nabla \nu \eqno(2.16)$$
which is a Clebsch form (Lamb (1945)).
It has been shown in paper I that the function $ \nu $ is single valued . The
value of $\nu$ is defined if we impose mass conservation , i.e. equation (2.2)
or
$$ \vec \nabla \cdot (\sigma \vec \nabla \nu ) = \vec \nabla \cdot [ \sigma ( -
\alpha \vec \nabla \beta + \vec \eta_c)] \eqno(2.17)$$
	This is an elliptic equation for $\nu$ whose solution depends on one arbitary
constant only because $ \sigma|_B = 0$. We chose $ \nu(\alpha = 0) = 0$ to make
$\nu$ unique.

Thus given a mass preserving trial configuration of a stationary Maclaurin flow
in moving coordinates, there exists a labelling $\alpha, \beta$ of the fluid
elements with $0 \leq \alpha \leq \alpha_B $ and $ 0 \leq \beta \leq 2 \pi $.
The labelling is isocirculational in the following sense: the circulation along
any contour $l$ defined by $ \alpha (\beta)$ depend on $ \alpha (\beta)$ only:
$$ \oint_l \vec W \cdot d \vec R = \oint_l \alpha(\beta) d \beta \eqno(2.18)$$
and not on $x^K(\alpha,\beta)$. Any deformed trial configuration in which fluid
elements labeled $\alpha,\beta$ with coordinates $x^K(\alpha,\beta)$ go to
points $ \tilde x^K(\alpha,\beta)$ will have the same circulation along the
contour defined by the same function $\alpha(\beta)$ in $\tilde x^k$
coordinates.

Reciprocally let $x^K(\alpha,\beta)$ define the positions of fluid elements in
a trial configuration labelled by $\alpha, \beta$ with the topology of fig.1,
with $\lambda(\alpha)$ given, $0 \leq \alpha \leq \alpha_B $, $ 0 \leq \beta
\leq 2 \pi $, and $\alpha_B$ given as well. Then the density of the
configuration $\sigma(x^K)$ is defined by (2.11) and  (2.14.b) as
$$ \sigma = \lambda( \alpha) { \partial (\alpha,\beta) \over \partial (x,y)}
\eqno(2.19)$$
and a $ \vec W (x^K)$ is defined by (2.16) in which $\nu$ is given by equation
(2.17) and $ \nu (\alpha = 0) = 0$. The flow so defined $\sigma(x^K), \vec W
(x^K)$ is both mass preserving and isocirculational. Notice that $\vec b$ and
$\Omega_c$ appear in $\vec W$ through $\nu$. The three numbers $\vec b$ and
$\Omega_c$, are, in general, related to the linear momentum $\vec P$ and the
angular momentum $J$ of the flow; $\vec P$ can always be take equal to zero:
$$ \vec P = \int_S \vec W \sigma d^2 x = 0 \eqno(2.20.a)$$
$$  J = \vec 1_z \cdot \int_S \vec R \times \vec W \sigma d^2 x = J_0
\eqno(2.20.b)$$
\bigskip

\noindent
{\it 2.3. Fixation of Coordinates }
\bigskip

	 For definiteness, the origin and orientation of coordinates have to be fixed.
For this we have only the topology of the flow. We shall fix the relative
position of our trial flow ( see fig. 1 ) in the following way:
	The point of zero relative velocity, where $\alpha = 0$, will be chosen as the
origin of coordinates, and the $x$ axis oriented along one of the principal
directions of the boundary contour $\alpha = \alpha_B$. This fixes in general
the coordinate
system of any trial configuration . In these
coordinates  $ \vec R(\alpha,\beta)$  must satisfy the following
 conditions:
$$ \vec R(0,\beta) = 0 \eqno(2.21.a)$$
$$ ({\partial x \over \partial \beta})_{\alpha = \alpha_B, \beta = 0} = 0
\eqno(2.21.b)$$
In addition we must have
$$ x(\alpha,0) \geq 0,\qquad y(\alpha,0) = 0 \eqno(2.21.c)$$
because we chose $\beta = 0$ to be the positive x- axis.
\bigskip

\noindent
{\it 2.4. The Energy Criteria for Stability}
\bigskip

	The following criteria of stability referes to what we called in
paper I the strong energy condition, in which
mass conservation always holds. Among all mass preserving and isocirculational
trial configurations
$ x^K (\alpha,\beta) $,  that satisfy (2.21), only those for which the energy
$$ E = \int \left[{1 \over 2} \vec W^{2} + \varepsilon(\sigma) + {1 \over 2}
\Phi \right] \sigma d^{3}x  \eqno(2.22)$$
under small displacements
$$ \vec R(\alpha,\beta) \rightarrow \vec R(\alpha,\beta)+ \vec \xi
(\alpha,\beta) \eqno(2.23)$$
is stationary, $ \Delta E = 0$, for given $\vec P$ and $J$ ($ \Delta \vec P =
\Delta J = 0 $), are real physical flows in that they satisfy Euler's equations
. Physical flows,  with { \it minimum} of energy are
stable with respect to small perturbations. The
condition of stability for flows with given linear and angular-momenta are thus
$$ \Delta^2 E >  0 \qquad {\rm at}  \qquad \Delta E = \Delta \vec P = \Delta J
= 0 \qquad {\rm with} \qquad  \Delta^2 \vec P = \Delta^2 J = 0 \eqno(2.24)$$
As emphasized in paper I, (2.24) is a necessary and sufficient condition of
stability if the gyroscopic term of the Lagrangian, the sum of those terms that
are linear in time derivatives, is equal to zero. The explicit expression for
the gyroscopic\footnote*{Here we deal only with the strong energy principle and
drop the sub index "strong" in (6.9) of paper I} term $\Delta^2 G$ obtained in
3-D flows holds here with minor changes. Thus, the necessary and sufficient
conditions of stability for flows with given linear and angular momentum are
equation (2.24) with
$$ \Delta^2 G = 0 \eqno(2.25)$$
\bigskip

\noindent
{\bf 3. Application to Maclaurin Disks }
\bigskip

\noindent
{\it 3.1. Stationary Configurations }

	The following is taken from Binney and Tremaine (1987) with some
changes of notations. In Maclaurin disks  the fluid rotates with uniform
angular velocity $ \vec   \Omega = \vec 1_z \Omega $ and the velocity field in
inertial coordinates is thus
$$ \vec W_0 = \vec \Omega \times \vec R = \Omega R^2 \vec \nabla \varphi
\eqno(3.1)$$
an index 0 referes to stationary quantities, $ \varphi $ is the polar angle,
$R$ the radial distance. Maclaurin disks appear the same in coordinates
rotating with uniform angular velocity $\Omega_c$. In such coordinates,
following equation (2.1), the relative velocity
$$ \vec U_0 = (\vec \Omega - \vec \Omega_c) \times \vec R \eqno(3.2a)$$
and
$$ \vec b_0 = 0 \eqno(3.2b)$$
The vorticity
$$\omega_0 = \vec 1_z \cdot rot \vec W_0 = 2 \Omega  \eqno(3.3)$$
The mass preserving density
$$ \sigma_0 = \sigma_C \sqrt {1 - {R^2 \over a^2}} \qquad R \le a,
\eqno(3.4.a)$$
$$ \sigma_0 = 0  \qquad \qquad R \ge a. \eqno(3.4.b)$$
Pressure and specific enthalpy are given by the equation of state
$$ P = {1 \over 2} \kappa \sigma^3 \eqno(3.5.a)$$
$$ h = {3 \over 2} \kappa \sigma^2 \eqno(3.5.b)$$
The gravitational field in the plane of the disk
$$ \Phi_0 = {1 \over 2} \Omega_0^2 R^2, \qquad \Omega_0^2 = {\pi^2 G \sigma_C
\over 2a} \eqno(3.6)$$
$\Omega_0$ is the angular velocity of a test particle on a circular orbit.
Euler's equations (2.7) relate $\Omega$ to $\Omega_0$, $\sigma_C$ and {\it a}:
$$ \Omega^2 = \Omega_0^2 - {3 \kappa \sigma_C^2 \over a^2} \eqno(3.7)$$
The mass $M$ and the angular momentum $J_0$ define $\sigma_C$ and $\Omega$ for
any given radius $a$:
$$ M = {2 \pi \over 3} \sigma_C a^2    \eqno(3.8.a) $$
$$   \vec P_0 = 0  \eqno(3.8.b) $$
$$ J_0 = { 4 \pi \over 15} \Omega \sigma_C a^4 = { M C_B \over 5 \pi}
\eqno(3.8.c) $$
where
$$C_B = 2 \pi a^2 \Omega \eqno(3.9)$$
is the circulation along the boundary.

\noindent
{\it 3.2. $(\alpha, \beta)$ Labelling and the Load $\lambda (\alpha)$ in
Maclaurin Disks.}

	The labels defined in section 2 are here as follows.  Lines of constant load $
{\sigma_0 \over \omega_0 } = const. $ are circles. The circulation along lines
of constant load
$$ C(R) = 2 \pi \Omega R^2 \eqno(3.10)$$
Thus, $\alpha(x^K)$ defined by equation (2.13) is proportional to $R^2$:
$$ 0 \le \alpha = {C \over 2 \pi} = \Omega R^2 \le \alpha_B = {C_B \over 2 \pi}
\eqno(3.11)$$
The load $\lambda( \alpha)$ is correspondingly
$$ \lambda (\alpha) = {\sigma_0 \over \omega_0} = \lambda_C \sqrt{1 - {\alpha
\over \alpha_B}} \eqno(3.12.a)$$
where, with $\sigma_C$ given in terms of $M$ by (3.8.a) and $\Omega$ related to
$C_B$ in (3.9), we have
$$ \lambda_C = {3M \over 2C_B} \eqno(3.12.b)$$
$\beta$ is defined by equation (2.14.b) which reduces here to
$$ \omega_0 = 2 \Omega  = { \partial (\alpha,\beta) \over \partial (x,y)}  = {1
\over R} { \partial (\alpha,\beta) \over \partial (R,\varphi)} = 2 \Omega
{\partial \beta \over \partial \varphi} \eqno(3.13)$$
If $\beta = 0$ is the $x$ axis, then
$$ \beta = \varphi \eqno(3.14)$$
Equations (3.1), (3.11) and (3.14) give
$$ \vec W = \alpha \vec \nabla \beta \eqno(3.15.a)$$
or
$$ \vec \nabla \nu_0 = 0 \eqno(3.15.b)$$
and with $\nu_0(\alpha =0) = 0$
$$ \nu_0 = 0 \eqno(3.16)$$

\noindent
{\it 3.3. Trial Configurations of Maclaurin Flows that Differ Slightly from a
Maclaurin Disk }

	Consider now mass preserving and isocirculational trial configurations $\vec R
(\alpha, \beta)$ that differ slightly from a Maclaurin disk $\vec R_0 (\alpha,
\beta)$
$$ \vec R(\alpha,\beta) = \vec R_0(\alpha,\beta) + \vec \xi(\alpha,\beta)
\eqno(3.17)$$
with
$$ |\vec \xi| \ll a \eqno(3.18)$$
Instead of $(\alpha, \beta)$, we may conveniently use $(R, \varphi)$ as labels.
A perturbed configuration is drawn in figure 1 with coordinates taken in such a
way that formulas (2.21) hold. $\vec R$ is given by equation (3.17), and $\vec
R$ as $\vec R_0$ satisfy both equation (2.21.a); we must then have also
$$ \vec \xi (R = 0,\varphi) = 0 \eqno(3.19.a)$$
Equations (2.21.b and c) will be satisfied by removing from $\vec \xi$ any
displacements that amounts to a change of the cut $\beta=0$ by an arbitary
infinitesimal function of $\alpha$ i.e. $ \delta \beta = \tau( \alpha)$ . In
terms of equation (3.11) and (3.23) [below  where $\delta$ is related to $ \vec
\xi $ taking index $K=2$], we may equivalently write that condition as
$$ R^2 \delta \beta = \vec 1_z \cdot \vec \xi \times \vec R = R^2 \tau( R^2)
\eqno(3.19.b)$$
 Equation (3.19.b) includes arbitrary rigid rotations, $\delta \beta = const.$

The density $\sigma$ of our {\it Maclaurin} flows change from $\sigma_0$ given
by equation (2.19) in which $\lambda( \alpha)$ is the load of Maclaurin disks
defined by equation (3.12) with the same $\lambda_C$ and the same $\alpha_B$.
These  conditions preserve  indeed vortex fluxes. Thus the change in $\sigma$
$$ \Delta \sigma = \sigma - \sigma_0= -\sigma_0 \vec \nabla \cdot \vec \xi
\eqno(3.20)$$
The velocity field $\vec W$ is given by (2.16) and therefore the change in
$\vec W$
$$\Delta \vec W = \vec W - \vec W_0 = -\vec \nabla \vec \xi \cdot \vec W_0 +
\vec \nabla \Delta \nu \eqno(3.21)$$
$\Delta \nu$ itself is defined by the perturbed mass-preserving equation
$\Delta [\vec \nabla \cdot (\sigma \vec U)] = 0$. To obtain  this equation it
is convenient to introduce the often used $\delta$-variations appearing already
in (3.19.b), associated with changes of the functions $\alpha^K (x^L)$ that
leave the lines of constant $\alpha^K$ fixed. Such changes are defined by
$$ \alpha^K (\vec R + \vec \xi) + \delta \alpha^K (\vec R + \vec \xi) =
\alpha^K (\vec R) \eqno(3.22)$$
To order 1,
$$ \delta \alpha^K = - \vec \xi \cdot \vec \nabla \alpha^K \eqno(3.23)$$
Just like $\Delta \alpha^K = 0$, now $\delta x^K = 0$. The greatest quality of
$\delta$-variations is that they commute with $x^K$ derivatives
$$ \delta \vec \nabla = \vec \nabla \delta \eqno(3.24)$$
One easily find that
$$ \delta = \Delta - \vec \xi \cdot \vec \nabla \eqno(3.25)$$
We shall move from $\Delta$ to $\delta$ variations and back at our convenience.
Thus $\Delta \sigma$ in equation (3.20) gives
$$ \delta \sigma = - \vec \nabla \cdot (\sigma_0 \vec \xi) \eqno(3.26)$$
and $\Delta \vec W$ in eq (3.21) gives
$$ \delta \vec W = -\vec \nabla \vec \xi \cdot \vec W + \vec \nabla \Delta \nu
- \vec \xi \cdot \vec \nabla \vec W \eqno(3.27)$$
Let us now go back to the equation for $\Delta \nu$ or $\delta \nu$. Following
equation (2.2), we have
$$ \delta [\vec \nabla \cdot (\sigma \vec U)]|_0 = \vec \nabla \cdot (\delta
\sigma \vec U_0  + \sigma_0 \delta \vec U) = 0 \eqno(3.28)$$
With equations (2.1) and (3.2.a), one can see that (3.28) is an elliptic
equation for $\delta \nu$:
$$ \vec \nabla \cdot (\sigma_0 \vec \nabla \delta \nu) = -\vec \nabla \cdot
[\delta \sigma \vec W_0 + \sigma_0 \delta (\alpha \vec \nabla \beta)] + \delta
\vec b \cdot \vec \nabla \sigma_0 - \vec 1_z \cdot \vec R \times \vec \nabla
\delta \sigma \Omega_c \eqno(3.29)$$
Equation (3.29)  with $\delta \nu (\alpha = 0) = 0$ defines $\delta \nu$
uniquely. $\delta \nu$ is a non homogeneous linear expression in  $\delta \vec
b$ and $\Omega_c$. The latter will now be retrieved from the conditions that
perturbed Maclaurin flows have same linear and angular momenta as the Maclaurin
disks.
\bigskip

\noindent
{\it 3.4. Global Constraints}

{\it a. Fixation of $\delta \vec b$}

	The equation (2.2) for mass conservation implies the following identity:
$$ \vec R \vec \nabla \cdot (\sigma \vec U) = \vec \nabla \cdot (\sigma \vec U
\vec R) - \sigma \vec U = 0 \eqno(3.30) $$
With (2.6) and (2.1), we have thus
$$ \vec P = M( \vec b + \vec \Omega_c \times \vec R_{CM}) \eqno(3.31.a)$$
where
$$ \vec R_{CM} \equiv {1 \over M} \int \vec R \sigma d^2 x \eqno(3.31.b)$$
defines the position of the center of mass. For our Maclaurin disks
$$ \vec P_0 = \vec b_0 = \vec R_{CM0} = 0 \eqno(3.32)$$
If we want to keep $ \vec P =0$, we must take
$$ \delta \vec b = -\vec \Omega_c \times \delta \vec R_{CM} \eqno(3.33)$$
This equation defines $\delta \vec b$, that appears in (3.29), in terms of
$\vec \xi$ and $ \Omega_c $.

{\it b. Fixation of $\Omega_c$ }

	$\Omega_c$ is arbitary in a Maclaurin disk but as soon as we destroy axial
symmetry, we can see the flow rotating. $\Omega_c$ will be defined by the
condition that Maclaurin flows have the same angular momentum as the Maclaurin
disks.
	Turning our attention to $J$, defined in (2.20.b), we obtain
$$ \Delta J = \vec 1_z \cdot \int (\vec \xi \times \vec W + \vec R \times
\Delta \vec W) \sigma d^2 x  \eqno(3.34)$$
At the stationary point (Maclaurin disks) the first term on the right hand side
$$ \int \vec \xi \times \vec W_0 \sigma_0 d^2 x = \int \vec \xi \times (\vec
\Omega \times \vec R) \sigma_0 d^2 x = \Omega \int \vec \xi \cdot \vec R
\sigma_0 d^2 x \eqno(3.35)$$
while, using (3.1) and (2.6), we can write the second term as
$$ \int \vec R \times \Delta \vec W|_0 \sigma_0 d^2 x = \int \vec R \times
(-\vec \nabla \vec \xi \cdot \vec W_0) \sigma_0 d^2 x = - \Omega \int \vec \xi
\cdot \vec R \sigma_0 d^2 x \eqno(3.36)$$
Since $(3.35) + (3.36) \equiv 0$, we see that
$$ \Delta J|_0 \equiv 0 \eqno(3.37)$$
Thus $ \Delta J|_0 = 0$ does not define $\Omega_c$. However $ \Delta^2 J|_0 =
0$ does. Indeed, following equation (3.34),
$$ \Delta^2 J|_0 = \vec 1_z \cdot \int (2 \vec \xi \times \Delta \vec W|_0 +
\vec R \times \Delta^2 \vec W|_0 ) \sigma_0 d^2 x = 0 \eqno(3.38)$$
which contains $\Delta \nu$ and $\Delta^2 \nu$ that depend  both linearly on
$\Omega_c$. Notice, however that $\Delta^2 \nu$ appears in the following
integral
$$ \int \vec R \times \vec \nabla \Delta^2 \nu|_0 \sigma_0 d^2 x = \int  \vec
\nabla \times (\vec R \Delta^2 \nu|_0 \sigma_0) d^2 x =  \vec 1_z \oint
\Delta^2 \nu|_0 \sigma_0 \vec R  \cdot d \vec R \eqno(3.39)$$
The latter is zero because $\Delta^2 \nu$ must be small and  $\sigma_0|_B = 0$.
Therefore $\Omega_c$ will only be defined in terms of $\Delta^2 J = 0$ through
$\Delta \nu$ that appears in $\Delta \vec W |_0$ and $\Delta^2 \vec W |_0$. The
explicit expression for $\Omega_c$ is somewhat complicated and will be
calculated in due place.

\noindent
{ \it 3.5. The Energy Principle}

	We may now apply our strongly constrained energy principle $ \Delta^2 E > 0$
to find stability limits in Maclaurin disks. With $\Delta \nu$ defined by
(3.29) in which $\Omega_c$ is given by (3.38) and $ \delta \vec b$ by (3.33),
we can  evaluate $\delta \vec W$ given in (3.27) and $\delta \sigma$ in (3.26).
Here is a differential equality convenient for calculating $\Delta^2 E$. It is
obtained from (2.22) in appendix A
$$ \Delta^2 E - \Omega \Delta^2 J = \delta^2 E - \Omega \delta^2 J = \int \{
\sigma (\delta \vec W)^2 + \delta \sigma \delta (h + \Phi) \}|_0 d^2 x
\eqno(3.40)$$
\bigskip

\noindent
{\bf 4. Spherical Harmonic Decompositions of the Perturbations}
\bigskip
\noindent
{\it 4.1. Spherical Harmonic Decompositions of $\vec \xi$ , $\delta \sigma$,
$\delta h$ and $\delta \Phi$}

{\it a. $\vec \xi$ defined by scalar functions}

	 It is a good thing to define $ \vec \xi$ in terms of two independant non
dimensional infinitesimal scalars $ \eta$ and $ \psi$ as follows:
$$ \vec \xi = a^2 [\vec \nabla \eta + rot \vec \psi], \qquad \vec \psi = \vec
1_z \psi \eqno(4.1)$$
 $\eta$ is thus defined in terms of $\vec \xi$ by
$$ \triangle \eta = {1 \over a^2} \vec \nabla \cdot \vec \xi \eqno(4.2)$$
Some boundary conditions are needed to make $\eta$ unique. If we take, say,
$$\eta|_B = 0 \eqno(4.3.a)$$
equation (4.2) has a unique solution. Equation (4.2) represents also the
condition of integrability of (4.1), considered as a set of two first order
differential equations for $\psi$, given $\vec \xi$ and $\eta$. Thus the $\psi$
equations  are integrable and define $\psi$ up to a constant. This constant we
shall fix by asking:
$$\psi|_C = 0 \eqno(4.3.b)$$.
\bigskip

\noindent
{\it b. Spherical Harmonic Decomposition of $\eta$ and $\psi$ }

 	We now decompose $\eta$ and $ \psi$  in normalised spherical harmonic
functions of $\chi$ and $\varphi$ with
$$ \chi ={\sigma_0  \over \sigma_C} = \sqrt {1 - {R^2 \over a^2}}, \qquad 0 \le
\chi \le 1. \eqno(4.4)$$
We have
$$ \eta = \sum_{l=m}^{\infty} \sum_{m=0}^{\infty} \eta_{lm} {\cal P}_l^m (\chi)
e^{im \varphi} + c.c. \eqno(4.5.a)$$
$$ \psi = \sum_{l=m}^{\infty} \sum_{m=0}^{\infty} \psi_{lm} {\cal P}_l^m (\chi)
e^{im \varphi} + c.c. \eqno(4.5.b)$$
 $\eta_{lm}$ and $\psi_{lm}$ are arbitary, infinitesimal, complex numbers
related by equations (4.3). Associated Legendre polynomials are defined in the
range $ -1 \le \chi \le 1 $, but $\eta$ and $ \psi$ are only defined in the
range $ 0 \le \chi \le 1 $. We may extend $\eta$, $\psi$ to the negative region
of $\chi$ in any way we want. However, these functions have bounded gradients
at $\chi = 0$. Indeed, since
$$ |\partial_R  \eta| < \infty,\qquad |\partial_R  \psi| < \infty
\eqno(4.6.a)$$
$$ |{1 \over R} \partial_{\varphi} \eta| < \infty, \qquad |{1 \over R}
\partial_{\varphi} \psi| < \infty \eqno(4.6.b)$$
hold and since
$$ \partial_R  = - {1 \over a \chi} \sqrt{1- \chi^2} \partial_{\chi},
\eqno(4.7)$$
then,
$$ \partial_{\chi} \eta|_{\chi=0} = \partial_{\chi}  \psi|_{\chi=0} = 0
.\eqno(4.8)$$
Equations (4.8) will be satisfied if we make symmetrical continuous extensions
$$ \eta(\chi) = \eta(-\chi), \eqno(4.9.a)$$
$$ \psi(\chi) = \psi(-\chi), \eqno(4.9.b)$$
whose expansions are given by (4.5) with $(l-m)$ even. The expansions in the
domain $ 0 \le \chi \le 1 $, of {\it arbitrary} continuous $\eta$ and $\psi$
that satisfy (4.6) everywhere are then given by (4.5) with $(l-m)$ even [Arfken
1985] . The boundary conditions (4.3.b) gives the following relations among the
$\psi_{l0}$'s:
$$ \sum_{l=0}^{\infty} \psi_{l0} = 0, \qquad l \ \ even \eqno(4.10.a)$$
while equation (4.3.a) gives a relation among $\eta_{lm}$'s: for every $m \ge
0$:
$$ \sum_{l=m}^{\infty}  \eta_{lm} {\cal P}_l^m (0) = 0, \qquad l-m \ \ even
\eqno(4.10.b)$$
Equation (4.10.a) defines, say, $\psi_{00}$ in terms of $\psi_{l \ne 0,0}$; and
 equations (4.10.b) define, say, $\eta_{ll}$ in terms of $\eta_{l,m \ne l}$.

\noindent
{\it c. Spherical Harmonic Decomposition of $\delta \sigma$, $\delta h$ and
$\delta \Phi$ }

 	 We can now calculate the spherical harmonic decomposition of $\delta \sigma$
with equation (3.26); from (3.4.a) and (4.4) we obtain
$$ \sigma_0 = \sigma_C \chi. \eqno(4.11)$$
With equation (4.1) $\delta \sigma$ can be written as:
$$ \delta \sigma = -a^2 \sigma_C [ \vec \nabla \cdot (\chi \vec \nabla \eta) +
\vec \nabla \chi \times \vec \nabla  \psi]. \eqno(4.12)$$
and from equation (4.12) and (4.5) we find that $ \delta \sigma $ has the
following expansion
$$ \delta \sigma = \sigma_C \sum_{l=m}^{\infty} \sum_{m=0}^{\infty} \sigma_{lm}
{{\cal P}_l^m (\chi) \over \chi} e^{im \varphi} + c.c. \qquad l-m \ even
\eqno(4.13.a)$$
in which
$$ \sigma_{lm} = [l(l+1) -m^2] \eta_{lm} + im \psi_{lm} \equiv k_{lm} \eta_{lm}
+ im \psi_{lm} \eqno(4.13.b)$$
With the expansion of $\delta \sigma$ we obtain directly through (3.5.b) the
spherical harmonic decomposition of $\delta h$:
$$ \delta h = 3 \kappa \sigma_C^2 \sum_{l=m}^{\infty} \sum_{m=0}^{\infty}
\sigma_{lm} {\cal P}_l^m (\chi) e^{im \varphi} + c.c. \qquad l-m \ even
\eqno(4.14)$$
and from the varied Poisson equation :
$$ \triangle \delta \Phi = 4 \pi G \delta \sigma \delta_D (z) \eqno(4.15)$$
where $\delta_D (z)$ is the Dirac function, we find the solution $\delta \Phi$
which has been given by Hunter (1963):
$$ \delta \Phi = 2 \Phi_B \sum_{l=m}^{\infty} \sum_{m=0}^{\infty} \Phi_{lm}
{\cal P}_l^m (\chi) e^{im \varphi} + c.c. \qquad l-m \ even \eqno(4.16.a)$$
in which
$$ \Phi_{lm} = - g_{lm} \sigma_{lm} \eqno(4.16.b)$$
where
$$  g_{lm} = {(l+m)!(l-m)! \over 2^{2l-1} [({l+m \over 2})!({l-m \over 2})!]^2}
< 1 \qquad l>0, \qquad l-m \ even \eqno(4.16.c)$$
\bigskip

\noindent
{\it 4.2. Spherical Harmonic Decomposition of $\delta \vec W$}

	Following equation (3.27) and equations (3.1) and (4.1), $\delta \vec W$ may
be written as a gradient plus a rotational, always a convenient form for vector
fields:
$$ \delta \vec W = a^2 \Omega [ \vec \nabla \zeta+ rot (2 \eta \vec 1_z)]
\eqno(4.17.a)$$
in which
$$ \zeta= {1 \over a^2 \Omega}(\Delta \nu - \vec \xi \cdot \vec W_0)- 2\psi
\eqno(4.17.b)$$
Notice that not only is $\zeta \ll 1$ , $ |\vec \nabla \zeta|$ must be small
with respect to $a^{-1}$. $\Delta \nu$ is defined by equation (3.29) and
therefore, with (3.29), we obtain an elliptic equation for $\zeta$:
$$ \vec \nabla \cdot ( \sigma_0 \vec \nabla \zeta) = {1 \over 2a^2 \Omega} [
-\vec \nabla \sigma_0 \cdot rot (2 \eta \vec \Omega) - (\vec \Omega - \vec
\Omega_c) \times \vec R \cdot \vec \nabla \delta \sigma + \vec \nabla \sigma_0
\cdot \delta \vec b]. \eqno(4.18)$$
We now expand $\zeta$ in spherical harmonics
$$  \zeta = \sum_{l=m}^{\infty} \sum_{m=0}^{\infty} \zeta_{lm} {\cal P}_l^m
(\chi) e^{im \varphi} + c.c.,\qquad  l-m \ even \eqno(4.19.a)$$
We also use equation (4.1)  and (4.5) to find the expansion of $\delta \vec b$.
With equation (4.5.a) for $\eta$ and equation (4.13) for $\delta \sigma$, we
obtain, after some straightforward but a little tedious calculations, that
$$ \zeta_{lm} = -{ im \over k_{lm} }[2 \eta_{lm} - (1 - {\Omega_c \over
\Omega})   \sigma_{lm}] , \qquad (l,m) \ne (1,1) \eqno(4.19.b)$$
and
$$ \zeta_{11} =  i[\sigma_{11} - 2  \eta_{11}] = -\psi_{11} -i \eta_{11}.
\eqno(4.19.c)$$
Equations (4.19) and (4.5.a) inserted into (4.17.a) give the spherical harmonic
expansion of $\delta \vec W$.
\bigskip

\noindent
{\it 4.3. Spherical Harmonic decomposition of $\Delta^2 E$ and $\Delta^2 J$}

	$\Delta^2 E$ is given by (3.40) with $\Delta^2 J = 0$ that defines $\Omega_c$.
Near the stationary point,
$$ \Delta^2 E = \int \{  \sigma (\delta \vec W)^2 + \delta \sigma \delta (h +
\Phi) \}|_0 d^2 x \eqno(4.20)$$
To obtain the spherical harmonic decomposition of $\Delta^2 E$, we replace
$\delta \sigma$ , $\delta h$, $\delta \Phi$ and $\delta \vec W$ by their
respective expansions (4.12) (4.14) (4.16) and (4.17), in $\Delta^2 E$
[detailed calculations are given in appendix B]. The result is as  the follows:
$$ \eqalignno{&\Delta^2 E=4\pi\sigma_C \Omega^2 a^4\sum_{(l=m)>1, m=0}^{\infty}
4  (k_{lm} - {m^2 \over k_{lm}}) |\eta_{lm}|^2 + [{ m^2 \over k_{lm}} (1 -
{\Omega_c \over \Omega})^2  - 1 + (1 - g_{lm}){\Omega_0^2 \over \Omega^2}]
|\sigma_{lm} |^2 \cr & &(4.21) \cr}$$
The parenthesis $(1 - {\Omega_c \over \Omega})$ is given by $ \Delta^2 J =0$
(the calculation is in appendix C):
$$ 1 - {\Omega_c \over \Omega} = {\sum (k_{lm}^2 - m^2) |\eta_{lm} -
{\sigma_{lm} \over k_{lm}}|^2 + {m^2 \over k_{lm}^2} |\sigma_{lm}|^2 \over \sum
{m^2 \over k_{lm}} |\sigma_{lm}|^2} \qquad (m \ne 0),(l,m \ne 1,1)  \eqno(4.22)
$$

\bigskip

\noindent
{\it 4.4. Fictitious Marginal Instabilities }

Direct inspection of $\Delta^2 E$ in equation (4.21) reveals that the following
modes are absent:
$\eta_{ll}$, $\sigma_{11} = \eta_{11} + i \psi_{11}$ and $\psi_{l0}$. Equation
(4.10.a) defines $\psi_{00}$, equations (4.10.b) define $\eta_{ll}$. The other
modes, $\psi_{l0} \ l>0$ and $\sigma_{11}$, must thus represent symmetries. It
is important to figure out what these are.

The absence of  $\sigma_{11} = \eta_{11} + i \psi_{11}$ in $\Delta^2 E$ is
related to $\Delta^2 E$'s invariance for uniform translations of the coordinate
axis. Translational invariance is disposed off by imposing equation (3.19.a)
whose spherical harmonic decomposition is
$$ \sum_l l(l+1) (\eta_{l1} + i \psi_{l1}) = 0 \eqno(4.23)$$
{}From which we see that $\sigma_{11} = \eta_{11} + i \psi_{11}$ can be
retrieved in terms of $\eta_{l1} + i \psi_{l1}$ for $l \ne 1$, Hence, the
origin of coordinates is redefined, in accordance with condition (2.21.a).

 The coefficients $\psi_{l0}$, $l > 0$ are defined by equation (3.19.b). If we
use (4.1) and the spherical harmonic decomposition (4.5), we obtain from
(3.19.b)
$$ {\vec 1_z \cdot \vec \xi \times \vec R \over R} = a^2  \sum_{l=0}^{\infty}
\psi_{l0} {d{\cal P}_l (\chi) \over dR}= R \tau(R^2) \eqno(4.24)$$
Terms dependent on $\varphi$ have been removed from the left because $\tau$ is
independant of $\varphi$. Equation (4.24) depend on $\psi_{l0}$, $l > 0$ only;
the latter can be retrieved explicitly  with two obvious integrations:
$$ \psi_{l0} = \int_0^1 {\cal P}_l (\chi) [\int_{\chi}^1 \tau(\chi') \chi'd
\chi'] d\chi \qquad l>0 \eqno(4.25)$$
The way to keep the cut $\beta=0$ fixed on the $x$ axis is simply to set
$\psi_{l0} = 0$ for $l>0$. Notice that $\psi_{20} = 0$ alone avoids rigid
rotations.
\bigskip

\noindent
{\bf 5. Necessary and Sufficient Conditions for Stability of Maclaurin Disks}
\bigskip

\noindent
{\it 5.1. The Gyroscopic Term}

Following section 2.4 the inequality $\Delta^2 E > 0$, calculable from (4.21)
with (4.22), gives sufficient conditions of stability. $\Delta^2 E $ has the
form $ { \Omega_0^2 \over \Omega^2} A^2-B > 0$. If $B < 0$, the disk is stable
for any ${ \Omega^2 \over \Omega_0^2}$. For instance, Maclaurin disks are
stable to any perturbation that keeps the same densities at the displaced
points $(\delta \sigma = 0)$. We are naturally interested in perturbations that
might upset stability and for which $B > 0$. If $B > 0$ then ${ \Omega^2 \over
\Omega_0^2}$ must satisfy the following inequality
$$ \eqalignno{& Q \equiv { \Omega^2 \over \Omega_0^2} < {\sum_{(l=m)>1,
m=0}^{\infty} (1 - g_{lm}) |\sigma_{lm} |^2 \over \sum_{(l=m)>1, m=0}^{\infty}
[1-{ m^2 \over k_{lm}} (1 - {\Omega_c \over \Omega})^2 ] |\sigma_{lm} |^2  - 4
(k_{lm} - {m^2 \over k_{lm}}) |\eta_{lm}|^2 } \cr & &(5.1) \cr}$$
 The condition (5.1) becomes sufficient and necessary when the gyroscopic term
for dynamical perturbations $\Delta^2 G$ is zero (see paper I). It is therefore
important to calculate $\Delta^2 G$. For dynamical perturbations, $\vec \xi$ is
a function of the time $t$ as well as of $\alpha,\beta$ . Gyroscopic terms are
those bilinear functionals of $\vec \xi$ and $\dot {\vec \xi} = {\partial \vec
\xi \over \partial t}$ which appear in the Lagrangian of the dynamically
perturbed equations. We have calculated $\Delta^2 G$ in paper I. For perturbed
Maclaurin Disks, $\Delta^2 G$ of paper I reduces here to a rather simple
expression [see appendix D]
$$  \Delta^2 G = 2 \int_{t_0}^t \int   \dot {\vec \xi} \cdot \Delta \vec W
\sigma_0 d^2 x dt \eqno(5.2)$$
We can make a spherical harmonic decomposition of $\Delta^2 G$, with $\vec \xi$
given by (4.1) and (4.5) and $\Delta \vec W$ given in appendix by (C.5) . The
only novelty in  these calculations is that $\eta_{lm}$ and $\psi_{lm}$  are
now functions of $t$. A straightforward subtitution of $\vec \xi$ and $\Delta
\vec W$ in equation (5.2) leads to the following expression for
$\Delta^2 G$:
$$ \eqalignno{ &\Delta^2 G = 4 i \pi \sigma_C \int_{t_0}^t dt \sum_{l,m \ne
1,1} m \Omega \{ ({(1 - {\Omega_c \over \Omega})
\over k_{lm}} - {1 \over m^2}) \dot \sigma_{lm}^* \sigma_{lm} + \cr
&(1 -{k_{lm}^2 \over m^2}) \dot \eta_{lm}^* \eta_{lm} - 2 ({1 \over k_{lm}} -
{k_{lm} \over m^2}) \dot \eta_{lm}^* \sigma_{lm} \} + complex \  conjugate
&(5.3) \cr} $$
in which $(1 - {\Omega_c \over \Omega})$ has to be replaced by the value given
in equation (4.22).{ \it Necessary and sufficient conditions of stability are
obtained from (5.1) when $\Delta^2 G = 0$}.
\bigskip

\noindent
{\it 5.2. Symmetric and Antisymmetric Single-Mode Perturbations }
\bigskip

For symmetrical and antisymmetrical modes, we take either real or imaginary
components of $\eta_{lm}$ and $\psi_{lm}$ in the spherical harmonic expansion,
so the Fourier expansion contains either $cos(m \varphi)$ or $sin(m \varphi)$.
For such perturbations, the written part of $\Delta^2 G$ is imaginary; adding
the complex conjugate makes thus  $\Delta^2 G = 0$. For one single mode
$(l,m)$, equation (5.1) reduces to:
$$ Q < Q_{lm} = { (1 - g_{lm}) \sigma_{lm}^2 \over  [1-{ m^2 \over k_{lm}} (1 -
{\Omega_c \over \Omega})^2 ] \sigma_{lm}^2  - 4  (k_{lm} - {m^2 \over k_{lm}})
\eta_{lm}^2 } \eqno(5.4)$$
in which
$$ 1 - {\Omega_c \over \Omega} = {(k_{lm}^2 - m^2) (\eta_{lm} - {\sigma_{lm}
\over k_{lm}})^2 + {m^2 \over k_{lm}^2} \sigma_{lm}^2 \over {m^2 \over k_{lm}}
\sigma_{lm}^2} \qquad (m \ne 0),(l,m \ne 1,1)  \eqno(5.5) $$
It is advatageous to introduce a single arbitary variable $z$ and a constant
$x$ to analyze the lower bound of $Q_{lm}$:
$$ z \equiv - k_{lm} { \eta_{lm} \over \sigma_{lm}}  \qquad -\infty < z <
\infty \eqno(5.6.a)$$
and
$$ x \equiv { m^2  \over k_{lm}^2} \qquad 0 \le x \le 1 . \eqno(5.6.b)$$
In term of $z$
$$ Q_{lm} = {(1-g_{lm}) \over P_4(z)} \eqno(5.7.a)$$
where $P_4(z)$ is a polynomial of order 4 in $z$:
$$P_4 (z) = 1 - {1 \over k_{lm}} \{({1 \over x} -1)[(1-x)z^4 +4(1-x)z^3 + 6z^2
+4z] + {1 \over x} \} \eqno(5.7.b)$$
$P_4(z)$ has one and only one real maximum for any $x$ (see figure 2).
We are only interested in values of $z$ for which $P_4(z)>0$. For $P_4(z)<0$,
$\Delta^2 E>0$ for any value of Q. The maximum of $P_4(z)$ is obtained for:
$$ z_{max}(x) = {x^{{1 \over 3}} \over \sqrt{1-x}} [(\sqrt{1-x}-1)^{1 \over 3}
+ (\sqrt{1-x}+1)^{1 \over 3}] - 1 \eqno(5.8.a)$$
for which
$$P_4 (z)_{max} = 1 - {1 \over k_{lm}} \{({1 \over x} -1)[(1-x)z_{max}^4
+4(1-x)z_{max}^3 + 6z_{max}^2 +4z_{max}] + {1 \over x} \} \equiv 1 - {y(x)
\over k_{lm}} \eqno(5.8.b)$$
The function $y(x)$ appears in figure 3. To any  pair of values $(l,m)$
corresponds a value $x$ defined by (5.6.b) and a point $(x,y)$ on the curve.
Following (5.8.a) and (5.4), we must thus  have
$$ Q<(Q_{lm})_{min} = {(1-g_{lm}) \over 1 - {y(x) \over k_{lm}}} \eqno(5.9)$$
The smallest minimum of $Q_{lm}$ is obtained for $(l,m)=(2,2)$ for which $
(Q_{22})_{min}={1 \over 2}$. Therefore the necessary and sufficient condition
of stability with respect to symmetric or antisymmetric single mode
perturbations of Maclaurin disks is
$$ Q < {1 \over 2} \eqno(5.10)$$
\bigskip

\noindent
{\it 5.3 Comments}
\bigskip
\noindent
a)  Binney and Tremaine have given the following dispersion relation for the
$\omega$ - modes of dynamical perturbations:
$$ \omega_r^3 - \omega_r \{ 4 \Omega^2 + k_{lm}[\Omega_0^2(1- g_{lm}) -
\Omega^2] \} +2m\Omega[\Omega_0^2(1- g_{lm}) - \Omega^2]=0, \qquad \omega_r =
\omega - m \Omega \eqno(5.11)$$
Stability holds if the non spurious $\omega_r$'s are real roots. Equation
(5.11) is a third order polynome. The condition for a polynome  of the form
$x^3 + a_2 x +a_3 = 0$ to have only real roots is given in standard handbooks
[for instance: Schaum's Mathematical Handbook (1968)]
$$ ({a_2 \over 3})^3+({a_3 \over 2})^2 < 0 \eqno(5.12)$$
Inequality (5.12) for equation  (5.11) is exactly  our inequality (5.9).
\bigskip
\noindent
b) For $l=m$ modes , $\Delta^2 G = 0$ for { \it asymmetrical} modes, that is,
for $\sigma_{ll}$ complex [  $\eta_{ll}$  does not appear in $\Delta^2 G$ nor
in $\Delta^2 E$] . However, the stability limit is still given by equation
 (5.9) with $l=m$, for which $y=1$ and $k_{lm} = l$:
$$ Q < Q_{ll} = {1 - g_{ll} \over 1 - {1 \over l }}, \qquad g_{ll} = {(2l)!
\over 2^{2l-1} l!^2} \eqno(5.13)$$
\bigskip
\noindent
c) For radial modes, $m=0$, $\Delta^2 E$ does not depend on $\Omega_c$,
$\Delta^2 J \equiv 0$ and $\Delta^2 G = 0$. Moreover, since [see equation
(4.13.b)]
$$ \sigma_{l0} = l(l+1) \eta_{l0} \eqno(5.14)$$
 all $l$-modes are decoupled, $\Delta^2 E$ is a sum of squares of independent
modes. In this case one obtains stability limits for non single modes. The most
unfavorable limit, however, is again that given by $Q_{l0}$ for which $y=4$ and
$k_{ll} = l(l+1)$:
$$ Q <  Q_{l0} = {1 - g_{l0} \over 1 - {4 \over l(l+1)}} \qquad (l \ge 2),
\qquad g_{l0} = {l!^2 \over 2^{2l-1} [({l \over 2})!]^4} \eqno(5.15)$$
\bigskip
\noindent
d) For coupled $l=m$ modes, $\Delta^2 G \ne 0$ and we obtain only sufficient
conditions of stability. In particular if we couple a pair of modes $l=m$ and
$l'=m'$ , sufficient conditions of stability go from $Q < Q_{ll}$ for $l'=l$ to
$$ Q < 1-g_{l'l'} \eqno(5.16)$$
for $l \gg l'$. The smallest values of (5.16) is ${1 \over 4}$ which is also
the secular limit of stability found by Binney and Tremaine (1987).

\bigskip

Joseph Katz's contribution to this work started during his stay at the National
Astronomical Observatory in Mitaka, whose hospitality he gratefully
acknowledged.
\vfill\eject

\noindent
{\bf References}

\refs
Arfken G. 1985 {\it Mathematical Methods for Physicists} Chap. 12 Academic
Press. P. 637 - 712.

Binney J. \& Tremaine S. 1987, {\it Galactic Dynamics} Chap. 5 Princeton
University Press.

Hunter C. 1963 {\it Month. Not. Roy. Astro. Soc.} {\bf 126} 23.

Katz J., Inagaki S. \& Yahalom A. 1993, {\it Publ.Astron.Soc.Japan} {\bf 45}
421.

Katz J. \& Lynden-Bell D. 1985 {\it Geophys. Astrophys. Fluid Dynamics}
{\bf 33} 1.

Lamb H. 1945 {\it Hydrodynamics} Dover Publications. P. 248

Lynden-Bell D. \& Katz J. 1981 {\it Proc. Roy. Soc. London} {\bf A 378}
179.

Spiegel M. 1968, {\it Mathematical Handbook} Macgraw-Hill, Schaum's outline
series p. 32.

Yahalom A. 1993 Ph.D. Thesis, Hebrew University of Jerusalem (Unpublished).

\endrefs
\vfill\eject
\noindent
{\bf  Appendix A: $ \Delta^2 E - \Omega \Delta^2 J$ Variational Identity
(3.40)}

We start from (2.22). Varying (2.22), using (2.8) for $\varepsilon$ and (2.9)
for $\Phi$,  and calculating at $\Delta E = 0$, we have
$$ \Delta^2 E = \int \{ (\Delta \vec W)^2 + \vec W \cdot \Delta^2 \vec W +
\Delta [\vec \xi \cdot \vec \nabla (h + \Phi)] \} |_0 \sigma_0 d^2 x
\eqno(A.1)$$
By substracting from $\Delta^2 E$, $\Omega \Delta^2 J$ where $\Delta^2 J$ is
given in (3.38), we obtain
$$ \Delta^2 E - \Omega \Delta^2 J = \int \{ (\Delta \vec W)^2  + \vec W \cdot
\Delta^2 \vec W - \vec \Omega \cdot (2 \vec \xi \times \Delta \vec W + \vec R
\times \Delta^2 \vec W) + \Delta [\vec \xi \cdot \vec \nabla (h + \Phi)] \}|_0
\sigma_0 d^2 x \eqno(A.2)$$
The $\Delta^2 \vec W$ drops out of equation (A.2) because of (3.1) . The  terms
containing $\Delta \vec W$ can be written as a difference of squares:
$$(\Delta \vec W)^2  - \vec \Omega \cdot (2 \vec \xi \times \Delta \vec W) =
(\Delta \vec W - \vec \Omega \times \vec \xi)^2 - (\vec \Omega \times \vec
\xi)^2 = (\Delta \vec W - \vec \xi \cdot \vec \nabla \vec W_0)^2 - \Omega^2
\vec \xi^2 = (\delta \vec W )^2 - \Omega^2 \vec \xi^2 \eqno(A.3)$$
In the last equality we have used (3.25). So equation  (A.2) with (A.3) can now
be written:
$$ \Delta^2 E - \Omega \Delta^2 J = \int \{ (\delta \vec W)^2 - \Omega^2 \vec
\xi^2 + \Delta [\vec \xi \cdot \vec \nabla (h + \Phi)] \}|_0 \sigma_0 d^2 x
\eqno(A.4)$$
With the following operator identity
$$ \Delta \vec \nabla  = \vec \nabla \Delta  - \vec \nabla \vec \xi \cdot \vec
\nabla  \eqno(A.5)$$
and with equation (3.25), we may rewrite the $h + \Phi$ term in (A.1) as
follows:
$$ \Delta [\vec \xi \cdot \vec \nabla (h + \Phi)] = \vec \xi \cdot \vec \nabla
\Delta (h + \Phi) - \vec \xi \cdot \vec \nabla \vec \xi \cdot \vec \nabla (h +
\Phi) = \vec \xi \cdot \vec \nabla \delta (h + \Phi) + \vec \xi \cdot \vec
\nabla  \vec \nabla (h + \Phi) \cdot \vec \xi \eqno(A.6)$$
But from Euler's equations (2.7) we see that
$$\vec \nabla (h + \Phi)|_0 = - \vec W_0 \cdot \vec \nabla \vec W_0
\eqno(A.7)$$
Subtituting (A.7) into (A.6) and $ \Delta [\vec \xi \cdot \vec \nabla (h +
\Phi)]$ back in (A.4), we find that
$\Omega^2 \vec \xi^2$ cancels out. Finally with $\delta \sigma$ given in
(3.26), and with some integration by parts, we obtain the following much
simpler form for $\Delta^2 E- \Omega \Delta^2 J$
which is that written in (3.40):
$$ \Delta^2 E - \Omega \Delta^2 J = \int \{  \sigma (\delta \vec W)^2 + \delta
\sigma \delta (h + \Phi) \}|_0 d^2 x \eqno(A.8)$$
\bigskip

\noindent
{\bf  Appendix B: Spherical Harmonic Decomposition of $\Delta^2 E$}

Let us start from (A.8) and take $\Delta^2 J = 0$. We shall set
$$ \Delta^2 E = \Delta^2 E_k + \Delta^2 E_p \eqno(B.1.a)$$
in which
$$ \Delta^2 E_k = \int  [\sigma (\delta \vec W)^2]|_0 d^2 x \eqno(B.1.b)$$
and
$$ \Delta^2 E_p = \int [\delta \sigma ( \delta h + \delta \Phi)]|_0 d^2 x .
\eqno(B.1.c)$$
Inserting (4.17.a) in (B.1.b) gives:
$$ \Delta^2 E_k =  a^4 \Omega^2 \int \{ (\vec \nabla \zeta)^2 + 2 \vec \nabla
\cdot [\zeta rot (2 \eta \vec 1_z)] + [rot (2 \eta \vec 1_z)]^2 \} \sigma_0 d^2
x   \eqno(B.2)$$
Using identity (2.6), we can integrate (B.2) by part, and with $\sigma_0|_B=0$,
obtain:
$$ \Delta^2 E_k = a^4 \Omega^2 \int [ - \zeta \vec \nabla \cdot ( \sigma_0 \vec
\nabla \zeta) - 2 \zeta \vec \nabla \sigma_0 \cdot rot (2 \eta \vec 1_z) - 4
\eta \vec \nabla \cdot ( \sigma_0 \vec \nabla \eta)] d^2 x.   \eqno(B.3)$$
The spherical harmonic decomposition of $\Delta^2 E_k$ is obtained by replacing
in equation (B.3), $\zeta$ by (4.19), $\sigma_0$ by $\sigma_C \chi$ (equation
(4.11)) and $\eta$ by (4.5.a). The result comes out as follows:
$$ \Delta^2 E_k = 4 \pi  \sigma_C \Omega^2 a^4 \sum_{l=m, m=0}^{\infty} [4
(k_{lm} - {m^2 \over k_{lm}}) |\eta_{lm}|^2 + { m^2 \over k_{lm}}(1 - {\Omega_c
\over \Omega})^2 |\sigma_{lm}|^2 ] \eqno(B.4)$$
Notice that these are already some of the terms of $\Delta^2 E$ given in (4.21)

The spherical harmonic decomposition of $\Delta^2 E_p$ follows by inserting in
(B.1.c) the respective expansions of $\delta \sigma$ in (4.13), $\delta h$ in
(4.14) and $\delta \Phi$ in (4.15). For $ \Delta^2 E_p$,  we obtain :
$$ \Delta^2 E_p = 2 \pi a^4 \Omega_0^2 \sum_{l=m}^{\infty} \sum_{m=0}^{\infty}
         \sigma_{lm}[\Phi^*_{lm}
	+ 3 \kappa \sigma_C \sigma^*_{lm}] + c.c. \eqno(B.5)$$
With (4.16.b) and (3.7), $\Delta^2 E_p$ can also be written
$$ \Delta^2 E_p =  4 \pi \sigma_C a^4 \Omega^2 \sum_{l=m, m=0}^{\infty} [-1 +
(1 - g_{lm}) {\Omega_0^2 \over \Omega^2}] |\sigma_{lm}|^2  \eqno(B.6)$$
The sum of $\Delta^2 E_k$ given by (B.4) and $\Delta^2 E_p$ of (B.6) is the
expression of $\Delta^2 E$ written in (4.21).
\bigskip

\noindent
{\bf  Appendix C: Spherical Harmonic Decomposition of $\Delta^2 J$}

Following (3.38):
$$ \Delta^2 J = \vec 1_z \cdot \int (2 \vec \xi \times \Delta \vec W +\vec R
\times \Delta^2 \vec W)|_0  \sigma_0 d^2 x. \eqno(C.1)$$
By varying  $\Delta \vec W$ in (3.21), we obtain $\Delta^2 \vec W$:
$$ \Delta^2 \vec W = -2 \vec \nabla \vec \xi \cdot \Delta\vec W + \vec \nabla
\Delta^2 \nu  \eqno(C.2) $$
Inserting this into (C.1) gives:
$$ \Delta^2 J = \vec 1_z \cdot \int (2 \vec \xi \times \Delta \vec W -2 \vec R
\times  \vec \nabla \vec \xi \cdot \Delta\vec W )|_0  \sigma_0 d^2 x.
\eqno(C.3)$$
Notice that $\vec \nabla \Delta^2 \nu $ do not contribute to (C.3) [see
equation (3.39)]. Following equation (3.25) we can write:
$$\Delta \vec W = \delta \vec W + \vec \xi \cdot \vec \nabla \vec W_0
\eqno(C.4)$$
Using (3.1) for $\vec W_0$, (4.1) for $\vec \xi$, and (4.17.a) for $\delta \vec
W$, we obtain for (C.4)
$$ \Delta \vec W = a^2 \Omega [\vec \nabla (\zeta + \psi) + rot ( \eta \vec
1_z)] \eqno(C.5) $$
Inserting (C.5) and (4.1) into (C.3) gives
$$  \eqalignno{ & \Delta^2 J = 2 a^4 \Omega \vec 1_z \cdot \int \{  [\vec
\nabla \eta + rot \vec \psi] \times [\vec \nabla (\zeta + \psi) + rot ( \eta
\vec 1_z)] \cr & - \vec R \times  \vec \nabla  [\vec \nabla \eta + rot \vec
\psi] \cdot [\vec \nabla (\zeta + \psi) + rot ( \eta \vec 1_z)] \} \sigma_0 d^2
x. &(C.6) \cr}$$
The spherical harmonic decomposition of $\Delta^2 J$ is obtained by replacing
in equation (C.6) $\zeta$ by (4.19), $\sigma_0$ by $\sigma_C \chi$ (equation
(4.11)) and $\eta$, $\psi$ by (4.5). Using also (4.13.b) for $\sigma_{lm}$, the
result comes out as follows:
$$\eqalignno{& \Delta^2 J = 4 \pi \sigma_C a^4 \Omega \sum [(k_{lm}^2 -m^2)
|\eta_{lm} - {\sigma_{lm} \over k_{lm}}|^2 + ({1 \over k_{lm}} - (1- {\Omega_c
\over \Omega} )){m^2 \over k_{lm}} |\sigma_{lm}|^2 + c.c.] \cr
 & = 8 \pi \sigma_C a^4 \Omega (\sum [(k_{lm}^2 -m^2) |\eta_{lm} - {\sigma_{lm}
\over k_{lm}}|^2 + {m^2\over k^2_{lm}}|\sigma_{lm}|^2 - (1- {\Omega_c \over
\Omega} ) \sum {m^2 \over k_{lm}} |\sigma_{lm}|^2) &(C.7) \cr}$$
and demanding $ \Delta^2 J =0$ we see that:
$$ 1 - {\Omega_c \over \Omega} = {\sum (k_{lm}^2 - m^2) |\eta_{lm} -
{\sigma_{lm} \over k_{lm}}|^2 + {m^2 \over k_{lm}^2} |\sigma_{lm}|^2 \over \sum
{m^2 \over k_{lm}} |\sigma_{lm}|^2} \qquad (m \ne 0),(l,m \ne 1,1)  \eqno(C.8)
$$
which is equation (4.22).
\bigskip

\noindent
{\bf  Appendix D: $\Delta^2 G$ For Maclaurin Disks}

Equation (6.9) of paper I reads as follows (we drop the index strong):
$$  \eqalignno{ &\Delta^2 G = 2 \int \int [\vec \nabla
 \Delta \nu_D \cdot \vec \nabla  \Delta \nu_S  - \dot {\vec \xi} \cdot \vec
\nabla \vec \xi \cdot \vec W_0] \sigma_0
 d^2 x dt \cr &- \int \int [\Delta \vec \eta_{cD} \cdot \vec \nabla  \Delta
\nu_S + \Delta \vec \eta_{cS} \cdot \vec \nabla  \Delta \nu_D
+ \Delta \vec \eta_{cD} \cdot \vec \nabla \vec \xi \cdot \vec W_0
\cr &+ (\Delta \vec \Omega_{cD} \times \vec W_0 + \vec \Omega_c \times \vec
\nabla  \Delta \nu_D) \cdot \vec \xi]|_0  \sigma_0 d^2 x dt,
&(D.1) \cr} $$
$\nu$ and $ \Delta \nu $ are defined by the time dependent equation of mass
conservation and it's first variation. $ \Delta \nu $ may be decomposed into a
"dynamical contribution" $ \Delta \nu_D$ which is zero when $ \dot {\vec \xi} =
0 $ and a "steady" part $ \Delta \nu_S$ which is the one we encountered in
perturbed  steady flows; $\Delta \nu_S$ is the $\Delta \nu = \delta \nu $ of
this paper which appears for the first time in (3.21); here, however:
$$ (\Delta \nu)_{Paper I} =  \Delta \nu_D +  \Delta \nu_S \eqno(D.2)$$
The equation of mass conservation is what defines equations for $ \Delta \nu_D$
and $\Delta \nu_S$:
$$  \vec \nabla \cdot ( \sigma_0  \vec \nabla \Delta \nu_D ) =  \vec \nabla
\cdot [\sigma_0 (\dot {\vec \xi} + \Delta \vec \eta_{cD})] \eqno(D.3.a)$$
$$  \vec \nabla \cdot ( \sigma_0  \vec \nabla \Delta \nu_S ) =  \vec \nabla
\cdot [\sigma_0 ( \vec U_0 \cdot \vec \nabla \vec \xi  + \vec \nabla \vec \xi
\cdot \vec U_0) + \sigma_0 \vec \nabla \vec \xi \cdot \vec \eta_c + \Delta \vec
\eta_{cS})] \eqno(D.3.b)$$
Similarly,  $ \vec \eta_c $ and $\Delta \vec \eta_c $(equation (2.1.b)) have a
dynamical and steady contribution  and we write
$$ \Delta \vec \eta_c = \Delta \vec \eta_{cD} + \Delta \vec \eta_{cS}
\eqno(D.4.a)$$
$$ \Delta \vec \eta_{cD} = \Delta \vec b_D + \Delta \vec \Omega_{cD} \times
\vec R \eqno(D.4.b)$$
$$ \Delta \vec \eta_{cS} = \Delta \vec b_S + \vec \Omega_{cS} \times \vec \xi +
\Delta \vec \Omega_{cS} \times \vec R \eqno(D.4.c)$$
The four last terms of (D.1) can be written as follows with the help of
$$\Delta \vec P_D = \int \vec \nabla \Delta \nu_D \sigma_0 d^2 x \eqno(D.5.a)$$
and
$$\Delta J_D = \vec 1_z \cdot \int  \vec R \times \vec \nabla \Delta \nu_D
\sigma_0 d^2 x \eqno(D.5.b)$$
and $\Delta \vec P_S $ and $\Delta  J_S $ which are the same as our $\Delta
\vec P $ leading to (3.33) and $\Delta  J $ given in (3.34); now using the
definitions in (D.4) we see that the four last terms of $\Delta^2 G$ (see D.1)
$$ \eqalignno{ & - \int \int [\Delta \vec \eta_{cS} \cdot \vec \nabla  \Delta
\nu_D + \Delta \vec \eta_{cD} \cdot \vec \nabla \vec \xi \cdot \vec W_0
\cr &+ (\Delta \vec \Omega_{cD} \times \vec W_0 + \vec \Omega_c \times \vec
\nabla  \Delta \nu_D) \cdot \vec \xi]|_0  \sigma_0 d^2 x dt,
&(D.6.a) \cr} $$
become:
$$ \eqalignno{ & - \int [\int   \Delta \vec \eta_{cD} \cdot \vec \nabla  \Delta
\nu_S \sigma_0 d^2 x] +\cr & \Delta \vec b_S \cdot \Delta \vec P_D + \Delta
\vec \Omega_{cS} \cdot \Delta \vec J_D - \Delta \vec b_D \cdot \Delta \vec P_S
- \Delta \vec \Omega_{cD} \cdot \Delta \vec J_S dt,
&(D.6.b) \cr} $$
Inserting (D.6.b) back into (D.1) gives thus:
$$  \eqalignno{ &\Delta^2 G = 2 \int \int [\vec \nabla
 \Delta \nu_D \cdot \vec \nabla  \Delta \nu_S  - \dot {\vec \xi} \cdot \vec
\nabla \vec \xi \cdot \vec W_0] \sigma_0
 d^2 x dt - \int \{ [\int  2 \Delta \vec \eta_{cD} \cdot \vec \nabla  \Delta
\nu_S \sigma_0 d^2 x] +\cr & \Delta \vec b_S \cdot \Delta \vec P_D + \Delta
\vec \Omega_{cS} \cdot \Delta \vec J_D - \Delta \vec b_D \cdot \Delta \vec P_S
- \Delta \vec \Omega_{cD} \cdot \Delta \vec J_S \} dt,
&(D.7) \cr} $$
Conservation of linear and angular momentum works now as follows.
$\Delta \vec J_S$ and $\Delta \vec J_D$ vanish identically due to the symmetry
of the problem. Define $\Delta \vec b_S$ and  $\Delta \vec b_D$ by setting
separately $\Delta \vec P_S = 0$ and $\Delta \vec P_D = 0$. With this,
$\Delta^2 G$ reduces to
$$  \Delta^2 G = 2 \int \int  [\vec \nabla
 \Delta \nu_D \cdot \vec \nabla  \Delta \nu_S - \dot {\vec \xi} \cdot \vec
\nabla \vec \xi \cdot \vec W_0 - \Delta \vec \eta_{cD} \cdot \vec \nabla
\delta \nu  ]  \sigma_0 d^2 x dt \eqno(D.8)$$
Integrating the first and the last term of (D.8) by parts: and replacing
finally $\Delta \nu_S$ by our present $\Delta \nu = \delta \nu$ defined in
(3.29), we obtain:
$$  \Delta^2 G = 2 \int \int  \{ \Delta \nu \vec \nabla \cdot [\sigma_0 (\Delta
\vec \eta_{cD} - \vec \nabla \Delta \nu_D)] - \dot {\vec \xi} \cdot \vec \nabla
\vec \xi \cdot \vec W_0 \sigma_0  \} d^2 x dt \eqno(D.9)$$
and with (D.3.a), (D.9) can also be written:
$$  \Delta^2 G = 2 \int \int [- \Delta \nu \vec \nabla \cdot (\sigma_0 \dot
{\vec \xi})- \dot {\vec \xi} \cdot \vec \nabla \vec \xi \cdot \vec W_0
\sigma_0] d^2 x dt \eqno(D.10)$$
And again,  integrating by parts the first term of the integrant we obtain:
$$  \Delta^2 G = 2 \int \int  \dot {\vec \xi} \cdot ( \vec \nabla \Delta \nu -
\vec \nabla \vec \xi \cdot \vec W_0) \sigma_0 d^2 x dt \eqno(D.11)$$
and finally using (3.21) in (D.12), we get the simple form of $\Delta^2 G$
written in equation (5.2):
$$  \Delta^2 G = 2 \int_{t_0}^t \int   \dot {\vec \xi} \cdot \Delta \vec W
\sigma_0 d^2 x dt \eqno(D.12)$$
\vfill\eject

{ \bf Figure Captions}

\noindent
Figure 1: Examples of a trial configuration. The drawing shows some $\alpha =
const$ and $\beta = const$ lines and the positioning of coordinate axis.

\noindent
Figure 2: This represents one $P_4 (z) > 0$ for $x = {1 \over 64}$ coresponding
to $(l,m) = (4,2)$ and a $z_{max} = -0.734$. The hatched part of this curve
corresponds to perturbations stable  for any $Q$.

\noindent
Figure 3:
(No Caption).
\bye